\documentclass[superscriptaddress,amssymb,amsmath,nofootinbib,onecolumn,aps,prd,longbibliography]{revtex4-1}
\usepackage[utf8]{inputenc}
\usepackage{mathtools}%para usar floor
\usepackage{hyperref}
\usepackage{slashed}
\hypersetup{colorlinks=true, urlcolor=blue,linkcolor=blue, citecolor=blue}
\usepackage{longtable}

\DeclarePairedDelimiter\floor{\lfloor}{\rfloor} %para facilitar uso do floor

\begin{document}
\title{The effect of boundary conditions on dimensionally reduced field-theoretical models at finite temperature}
\author{E. Cavalcanti}
\email[]{erich@cbpf.br}
\affiliation{Centro Brasileiro de Pesquisas F\'{\i}sicas/MCTI, 22290-180 Rio de Janeiro, RJ, Brazil}
\author{C.A. Linhares}
\email[]{linharescesar@gmail.com}
\affiliation{Instituto de F\'{\i}sica, Universidade do Estado do Rio de Janeiro, 20559-900 Rio de Janeiro, RJ, Brazil}
\author{J. A. Louren\c{c}o}
\email[]{jose.lourenco@ufes.br}
\affiliation{Departamento de Ci\^encias Naturais, Universidade Federal do Esp\'{i}rito Santo, 29932-540 Campus S\~ao Mateus, ES, Brazil}
\author{A. P. C. Malbouisson}
\email[]{adolfo@cbpf.br}
\affiliation{Centro Brasileiro de Pesquisas F\'{\i}sicas/MCTI, 22290-180 Rio de Janeiro, RJ, Brazil}

\begin{abstract}
Here we understand \textit{dimensional reduction} as a procedure to obtain an effective model in $D-1$ dimensions that is related to the original model in $D$ dimensions. %If a model in $D-1$ dimensions is not attainable we say that the dimensional reduction is not allowed. If the model in $D-1$ dimensions is exactly what we would find if we assumed from the beginning a reduced action, then the dimensional reduction is allowed and proper. There can also be a somewhat improper dimensional reduction where the model achieved in $D-1$ is not related with a reduced action of the original model.
To explore this concept we use both a self-interacting fermionic model and self-interacting bosonic model. Furthermore, in both cases, we consider different boundary conditions in space: periodic, antiperiodic, Dirichlet and Neumann.
For bosonic fields, we get the so defined dimensional reduction. Taking the simple example of a quartic interaction, we obtain that the boundary condition (periodic, Dirichlet, Neumann) influence the new coupling of the reduced model.
For fermionic fields, we get the curious result that the model obtained reducing from $D$ dimensions to $D-1$ dimensions is distinguishable from taking into account a fermionic field originally in $D-1$ dimensions. 
%This result seems to be related to the idea that scalar fields behave similarly in any dimension $D$ while fermionic fields are intimately related to the original dimension of the model.
Moreover, when one considers antiperiodic boundary condition in space (both for bosons or fermions) it is found that the dimensional reduction is not allowed.% This might be a topological property.  
\end{abstract}

\maketitle
    
%%%%%%%%%%%%%%%%%%%%%%%%%%%%%%%%%%%%%%%%%%%%%%%%%%%%%%%%%%%%%%%
%%%%%%%%%%%%%%%%%%%%%%%%%%%%%%%%%%%%%%%%%%%%%%%%%%%%%%%%%%%%%%%
%%%%%%%%%%%%%%%%%%%%%%%%%%%%%%%%%%%%%%%%%%%%%%%%%%%%%%%%%%%%%%%

\section{Introduction}
The construction and use of quantum field theoretical models at dimensions different from the usual space-time in $D=3+1$ are usual in the literature~\cite{Kaluza:1921tu,AliKhan:2001wr,Chakraverty:2002qk,Agashe:2004ay,Karch:2006pv,Blanke:2008yr,Fosco:2004cn,DaRold:2005mxj,Panico:2005dh,Panico:2006em,tHooft:1974pnl,Hands:2015qha,Rosenstein:1990nm,Hands:1992be,Rosenstein:1988dj,Klimenko:1991he,Klimenko:1992ch}. Its first appearance seems to be in the construction of the Kaluza five-dimensional theory~\cite{Kaluza:1921tu} that intended to unify gravity and electromagnetism. Since then, models and theories in $D\neq 4$ have been used in many different situations:
\begin{itemize}
    \item Phenomenology in particle physics considering extra dimensions ~\cite{AliKhan:2001wr,Chakraverty:2002qk,Agashe:2004ay,Karch:2006pv,Blanke:2008yr,Fosco:2004cn,DaRold:2005mxj,Panico:2005dh,Panico:2006em};
    \item Field theories in $D<4$ ~\cite{tHooft:1974pnl,Hands:2015qha,Rosenstein:1990nm,Hands:1992be,Rosenstein:1988dj,Klimenko:1991he,Klimenko:1992ch};
    \item Superstring theory~\cite{Horava:1996ma,Schwarz:1982jn,Aharony:1999ti}.
\end{itemize} 

In the context of finite-temperature field theory, it is understood that the regime of very high temperatures is associated with a dimensional reduction of the model. For scalar fields, it is possible to obtain an effective model in dimension $D-1$ that has a temperature-dependent coupling. This effective model is related to the original theory in $D$ dimensions when the temperature is very high~\cite{Appelquist:1981vg,Landsman:1989be,ZinnJustin:2002ru,MeyerOrtmanns:2007qp}. One of the uses of the thermal dimensional reduction is to investigate aspects of Hot QCD~\cite{Nadkarni:1982kb,Braaten:1995jr,Bialas:2000ev,Zhang:2011aa,Bazavov:2014pvz}. 

When we consider a system with restriction in one spatial direction, the discussion of dimensional reduction is renewed. For example, in the context of low-dimensional field theories ($D\le 4$), we can take into account the study of films and surfaces. Let us consider two physical systems: $(A)$ a film with thickness $L$ subjected to a thermal bath with temperature $T = 1/\beta$; $(B)$ a surface (planar system) subjected to the same temperature $T$. We call a \textit{dimensional reduction} the possibility that the model of the system $(A)$ becomes or brings information about a planar model - like the one of case $(B)$ - if we consider the limiting process to take the length to zero: $L\rightarrow 0$.

If we generalize this problem to an arbitrary number of dimensions we can ask ourselves whether there is a relationship between a model in $D$ dimensions and a model in $D-1$ dimensions; this is the major objective in the present study. 

It is a known theoretical result confirmed by experiments that for both bosonic and fermionic systems that undergo a phase transition, and are spatially limited, \emph{there is a minimum size below which there is no phase transition}~\cite{Khanna:2009zz,Khanna:2014qqa,Linhares:2006my}. This seems to indicate that for systems where at least one of the dimensions is restricted to a compact finite size with a compactification length $L$ a strict dimensional reduction is not allowed - at least in the context of phase transitions. Recently, in the context of phase transitions, it has been obtained that the minimal size of the system depends on the boundary conditions imposed on the spatial restriction. This analysis was done both for bosonic and fermionic models and a quasiperiodic boundary condition was applied which interpolates between the periodic and antiperiodic boundary conditions~\cite{Cavalcanti:2017wnm}.

We have previously found \cite{Cavalcanti:2018pgi} that for bosonic fields at the 1-loop level the so-called dimensional reduction is obtained when one considers periodic boundary condition in space. In this article, we extend this analysis so that we consider a few more boundary conditions: Dirichlet, Neumann and antiperiodic. Another step is to take into account purely fermionic models, so we can compare them with the bosonic situation. In the context of a thermal dimensional reduction, it is known that dimensional reduction happens for bosonic models~\cite{Appelquist:1981vg,Landsman:1989be,ZinnJustin:2002ru,MeyerOrtmanns:2007qp}. However, for fermionic fields, it seems that a model in $D$ dimensions is not related to a model originally built in $D-1$ dimensions~\cite{Huang:1995um,Ospedal:2017ubh}.   

%%%%%%%%%%%%%%%%%%%%%%%%%%%%%%%%%%%%%%%%%%%%%%%%%%%%%%%%%%%%%%%
%%%%%%%%%%%%%%%%%%%%%%%%%%%%%%%%%%%%%%%%%%%%%%%%%%%%%%%%%%%%%%%
%%%%%%%%%%%%%%%%%%%%%%%%%%%%%%%%%%%%%%%%%%%%%%%%%%%%%%%%%%%%%%%

\section{Generic model and boundary conditions}

Our aim is to discuss field theoretical models with self-interaction terms. In this way we avoid for the moment the combinatorics of many-particle models to focus on the effects of boundary conditions. The basic ingredient to discuss field theories in $D$ dimensions at one-loop level is the one-loop Feynman amplitude. In the scenario of a scalar field theory the amplitude $\mathcal{I}$ with $\rho$ propagators and zero external momenta is
\begin{equation*}
\mathcal{I}_\rho^D (M) = \int \frac{d^D p}{(2\pi)^D} \frac{1}{(p^2 + M^2)^\rho}. \label{Eq:I_original}
\end{equation*}
\noindent Where $M$ is the mass of the scalar field. The $D$-dimensional integral becomes an integral-sum after we introduce boundary conditions on $d<D$ coordinates. The compactification of the imaginary time introduces the inverse temperature $\beta = 1/T$ and the compactification of the spatial directions introduces some finite-lengths $L_i$. The boundary condition on the imaginary time must be periodic ($a_0=0$) for bosons or antiperiodic ($a_0=1$). However, there is freedom regarding the boundary condition imposed on the spatial direction. In the context of quantum field theories at toroidal topologies it has been discussed the use of periodic and antiperiodic boundary conditions~\cite{Khanna:2014qqa,Khanna:2009zz}, its extension to quasiperiodic boundary conditions~\cite{Cavalcanti:2017wnm} and also the use of Dirichlet and Neumann boundary conditions~\cite{Fucci:2017weg}. We consider a scenario with $d=2$ compactifications, after computing the remaining $D-2$ integrals using dimensional regularization we obtain that the one-loop Feynman amplitude for each boundary condition (\textit{b.c.}) is
\begin{equation}
\mathcal{I}_\rho^{D,2} (M;\beta,a_0,L|b.c.) = 
\frac{\Gamma\left(\rho - \frac{D}{2} +1 \right)}{(4\pi)^{\frac{D}{2}-1}\Gamma(\rho)\beta L} \sum_{n_0 \in \mathbb{Z}, n_1 \in \mathcal{M}} \left[M^2 + \left(\frac{2 \pi n_0}{\beta} + \frac{a_0 \pi}{\beta}\right)^2 + \omega_{n_1}^2 \right].  \label{Eq:I_compactified}
\end{equation}
\noindent Where the domain $\mathcal{M}$ of the sum over the frequencies $\omega_{n_1}$ is given in Table~\ref{Tab:BC} for each boundary condition
\begin{table}[!h]
    \begin{ruledtabular}
        \begin{tabular}{ccc}
            Boundary Condition (b.c.)    &    $\mathcal{M}$    & $\omega_{n_1}$\\ \hline
            Periodic ($\mathcal{P}$)                &    $\mathbb{Z}$        & $2 \pi n_1 / L$\\
            Antiperiodic ($\mathcal{A}$)            &    $\mathbb{Z}$    & $(2 n_1 + 1) \pi / L$\\
            Dirichlet ($\mathcal{D}$)                &    $\mathbb{N}^+$    & $\pi n_1 / L$\\
            Neumann ($\mathcal{N}$)                    &    $\mathbb{N}$    & $\pi n_1 / L$
        \end{tabular}
    \end{ruledtabular}
    \caption{Frequencies and domain of sum for each boundary condition in space.}
    \label{Tab:BC}
\end{table}

Although we start with a Feynman amplitude for a scalar field, Eq.\eqref{Eq:I_original}, it can be shown that the one-loop Feynman amplitude of $\mu$ fermionic propagators can be written as a combination of scalar one-loop Feynman amplitudes. We take into account a four-fermion coupling given by $a + b \gamma_S$, where $\gamma_S$ represents the chiral matrix. The one-loop Feynman amplitude in this scenario is
\begin{equation}
\mathcal{J}_\mu^D(M) = \text{tr}\int \frac{d^D p}{(2\pi)^D} \left(\frac{a+b\gamma_S}{i \slashed{p} + M}\right)^\mu.
\end{equation}
\noindent The relation between $\mathcal{J}_\mu^D$ and $\mathcal{I}_\rho^D$ is obtained in the Appendix~\ref{Ap:Fermion_Boson} and reads
\begin{multline}
\frac{1}{d_\gamma} \mathcal{J}_\mu^{D,d}  =
a^\mu \sum_{k=0}^{\floor{\frac{\mu}{2}}} \sum_{j=0}^{k} \binom{\mu}{2k} \binom{k}{j} M^{\mu-2j} (-1)^j \mathcal{I}^{D,d}_{\mu-j} (M)
+ b^\mu \left(\mu - 2 \floor{\mu/2}\right) \mathcal{I}^{D,d}_{\mu/2} (M)
\\
+ \sum_{k=1}^{\floor{\frac{\mu-1}{2}}} a^{\mu-2k} b^{2k} \sum_{j=k}^{\floor{\frac{\mu}{2}}} \frac{j! (\mu-j-1)!}{(j-k)!k!(\mu-k-j)!(k-j)!} \sum_{\ell=0}^{\floor{\frac{\mu}{2}-j}} \binom{\mu-2j}{2\ell}  \sum_{n=0}^{\ell} (-1)^n \binom{\ell}{n} M^{\mu-2j-2n} \mathcal{I}^{D,d}_{\mu-j-n} (M)
\\
+ \sum_{k=1}^{\floor{\frac{\mu-1}{2}}} a^{\mu-2k} b^{2k} \sum_{j=\floor{\frac{\mu}{2}}+1}^{\mu-k} \frac{j! (\mu-j-1)!}{(j-k)!k!(\mu-k-j)!(k-j)!} \sum_{\ell=0}^{\floor{\frac{\mu}{2}-j}} \binom{2j-\mu}{2\ell} \sum_{n=0}^{\ell} \binom{\ell}{n} M^{2j-\mu-2n} \mathcal{I}^{D,d}_{j-n} (M). \label{Eq:RelationJ_I}
\end{multline}
\noindent It holds independently of the number of compactified dimensions $d$. This means that the fermionic scenario is a combination of the relation given by Eq.~\eqref{Eq:RelationJ_I} and the expression of Eq.\eqref{Eq:I_compactified} considering antiperiodic boundary condition in the imaginary time ($a_0 = 1$). Therefore, in the analysis that follows, the bosonic behavior is studied by investigating Eq.\eqref{Eq:I_compactified} with $a_0=0$ and the fermionic behavior is studied by investigating Eq.\eqref{Eq:I_compactified} with $a_0=1$.

Notice that we can express both the cases of Dirichlet and Neumann boundary conditions in terms of the function with periodic boundary condition in space and a reduced function with just a thermal compactification.
\begin{align}
\mathcal{I}^{D,2}_\rho (M;\beta,a_0;L|\mathcal{D}) = \frac{1}{2}\mathcal{I}^{D,2}_\rho (M;\beta,a_0;2L|\mathcal{P}) - \frac{1}{2L} \mathcal{I}^{D-1,1}_\rho (M;\beta,a_0)
\label{Eq:FuncI_Dirichlet}
\\
\mathcal{I}^{D,2}_\rho (M;\beta,a_0;L|\mathcal{N}) = \frac{1}{2}\mathcal{I}^{D,2}_\rho (M;\beta,a_0;2L|\mathcal{P}) + \frac{1}{2L} \mathcal{I}^{D-1,1}_\rho (M;\beta,a_0)
\label{Eq:FuncI_Neumann}
\end{align}

Therefore, we only need to analyze the cases of periodic and antiperiodic boundary conditions in space. For both periodic and antiperiodic boundary conditions in space the remaining infinite sum in Eq.~\eqref{Eq:I_compactified} can be identified as an Epstein-Hurwitz zeta function~\cite{Elizalde:2012zza}. After an analytic continuation this leads to the sum over modified Bessel functions of the second kind $K_\nu(x)$; see Refs.~\cite{Khanna:2014qqa,Khanna:2009zz}. Using for convenience that $\nu = D/2-\rho$, the amplitude $\mathcal{I}^{D,2}_\rho$ reads
\begin{equation}
\mathcal{I}^{D,2}_\rho (M;\beta,a_0;L) = \frac{(M^2)^{\nu}\Gamma(-\nu)}{(4\pi)^{\frac{D}{2}}\Gamma(\rho)}
+ \frac{\mathcal{W}_{\nu} (M;\beta,a_0;L)}{(2\pi)^{\frac{D}{2}}2^{\rho-2}\Gamma(\rho)},
\end{equation}
\noindent where, for periodic boundary conditions in space ($\mathcal{P}$), the function $\mathcal{W}_\nu$ is
\begin{multline}
\mathcal{W}_\nu (M;\beta,a_0;L|\mathcal{P}) = \sum_{n=1}^{\infty} \cos(n\pi a_0)\left(\frac{M}{n \beta}\right)^\nu K_\nu (n \beta M) + \sum_{n=1}^{\infty} \left(\frac{M}{n L}\right)^\nu K_\nu (n L M) \\+ 2 \sum_{n_0,n_1=1}^{\infty} \cos(n_0\pi a_0) \left(\frac{M}{\sqrt{n_0^2 \beta^2 + n_1^2 L^2}}\right)^\nu K_\nu \left(M \sqrt{n_0^2 \beta^2 + n_1^2 L^2} \right),
\label{Eq:FuncW_Periodic_Initial}
\end{multline}
\noindent and, for antiperiodic boundary conditions in space ($\mathcal{A}$), the function $\mathcal{W}_\nu$ is
\begin{multline}
\mathcal{W}_\nu (M;\beta,a_0;L|\mathcal{A}) = \sum_{n=1}^{\infty} \cos(n\pi a_0)\left(\frac{M}{n \beta}\right)^\nu K_\nu (n \beta M) + \sum_{n=1}^{\infty} (-1)^n\left(\frac{M}{n L}\right)^\nu K_\nu (n L M) \\+ 2 \sum_{n_0,n_1=1}^{\infty} \cos(n_0\pi a_0)(-1)^{n_1} \left(\frac{M}{\sqrt{n_0^2 \beta^2 + n_1^2 L^2}}\right)^\nu K_\nu \left(M \sqrt{n_0^2 \beta^2 + n_1^2 L^2} \right).
\label{Eq:FuncW_AntiPeriodic_Initial}
\end{multline}

Notice that with the above equations one fully determines the behavior at one loop level for finite $\beta$ and finite $L$ both for bosonic and fermionic models in $D$ dimensions with the prescribed boundary conditions. In the following sections, we organize and apply the expressions for each situation under interest. 

%%%%%%%%%%%%%%%%%%%%%%%%%%%%%%%%%%%%%%%%%%%%%%%%%%%%%%%%%%%%%%%
%%%%%%%%%%%%%%%%%%%%%%%%%%%%%%%%%%%%%%%%%%%%%%%%%%%%%%%%%%%%%%%
%%%%%%%%%%%%%%%%%%%%%%%%%%%%%%%%%%%%%%%%%%%%%%%%%%%%%%%%%%%%%%%

\section{Dimensional reduction~\label{Sec:DimRed}}

In this section,  let us clarify the discussion of dimensional reduction. There are two main paths to obtain a dimensionally reduced field-theoretical model.

The first path is to take the original lagrangian in $D$ dimensions, reduce it to $D-1$ dimensions and then quantize it. This path ignores possible boundary conditions imposed on the removed dimension. The quantization is here understood as the computation of the correction given by the one loop Feynman amplitudes. If we are dealing with a model onde with one self-interacting bosonic field, the Feynman amplitude for the dimensionally reduced model in $D-1$ with one compactification corresponding to the inverse temperature $\beta = 1/T$ reads
\begin{equation}
\mathcal{I}^{D-1,1}_\rho (M;\beta,a_0=0) = \frac{\left(M^2/2\right)^\nu}{(2\pi)^\frac{D}{2}2^{\rho-2}\Gamma(\rho)}    
\left[
\frac{\pi}{M^2\beta} \Gamma(1-\nu) 
+ \frac{\sqrt{\pi}}{M} \sum_{k=0}^\infty \frac{(-1)^k}{\Gamma(k+1)} \Gamma\left(\nu-k-\frac{1}{2}\right) \zeta(2\nu-2k-1) \left(\frac{M \beta}{2}\right)^{-2\nu+2k+1}
\right].
\label{Eq:Id1_boson}
\end{equation}
\noindent On the other hand, for a model describing a self-interacting fermionic field, the Feynman amplitude $\mathcal{J}^{D-1,1}_\rho$ is related to $\mathcal{I}^{D-1,1}_\rho (M;\beta,a_0=1)$ through the relation given by Eq.~\eqref{Eq:RelationJ_I}, and the function $\mathcal{I}^{D-1,1}_\rho (M;\beta,a_0=1)$ reads
\begin{equation}
\mathcal{I}^{D-1,1}_\rho (M;\beta,a_0=1) = \mathcal{F}^D_\rho(M,\beta;c_1=0,c_2=1/2),
\label{Eq:Id1_fermion}
\end{equation}
\noindent where, for future convenience, the function $\mathcal{F}^D_\rho(M,\beta;c_1,c_2)$ is defined as
\begin{equation}
\mathcal{F}^D_\rho(M,\beta;c_1,c_2) = \frac{\left(M^2/2\right)^\nu}{(2\pi)^\frac{D}{2}2^{\rho-2}\Gamma(\rho)} \frac{\beta}{2\pi} \sum_{k=1}^{\infty} \frac{(-1)^k}{\Gamma(k+1)} \Gamma(k+1-\nu) \zeta(2k+2-2\nu) \left(\frac{M \beta}{2\pi}\right)^{2k-2\nu} (-1 + c_1 2^{2(k-\nu)} + c_2 2^{-2(k-\nu)} ).\label{Eq:F_function}
\end{equation}

We compare this first path with a different procedure to obtain a dimensionally reduced field theoretical model. In this second path, we take a quantized version of the model in $D$ dimensions and force the reduction taking the limit $L\rightarrow 0$. To explore this we need to evaluate $\mathcal{I}^{D,2}_\rho$ at a very small length $L$. We proceed as in Ref.~\cite{Cavalcanti:2018pgi} and use a integral representation of $K_\nu$ in the complex plane,
\begin{equation}
K_\nu(X) = \frac{1}{4\pi i} \int_{c-i\infty}^{c+i\infty} dt\; \Gamma(t) \Gamma(t-\nu) \left(\frac{X}{2}\right)^{\nu-2t}. \label{Eq:K_complex}
\end{equation}
\noindent To allow the interchange of the integral and the sum the value of $c$ must be chosen in such a way that there is no pole located to the right of it~\cite{Fucci:2017weg}. After using this integral representation we can compute the infinite sums and study the poles. It produces a tedious algebraic manipulation for each of the situations under interested and the main results are exhibited in the following subsections. Of course, this path splits into different ones as the choice of the boundary condition in the spatial direction might influence the result. Before investigating in further details the behavior as $L\rightarrow0$, let us reinforce that the investigation of the dimensional reduction comes from the comparison of both paths. This comparison may produce three different outcomes. 
\begin{itemize}
    \item At first, there might be a well-defined dimensional reduction, meaning that there is a relationship as 
    \begin{equation}
    s(L) \mathcal{I}^{D,2}_\rho (M;\beta,a_0;L|\text{b.c.}) \Bigg|_{L \rightarrow 0} = \mathcal{I}^{D-1,1}_\rho (M;\beta,a_0) + \{??\},
    \end{equation}
    \noindent where $s(L)$ is some \textit{scale} function that only depends on the finite length $L$, and we allow the presence of some residual terms.
    \item A second possibility is that the original model does not produce any relevant behavior as $L\rightarrow0$, and then the procedure of dimensional reduced is ill-defined and not allowed.
    \item A final possibility that could arise is that a dimensionally reduced model is achieved, but it does not correspond to the expected one.
    \begin{equation}
    s(L) \mathcal{I}^{D,2}_\rho (M;\beta,a_0;L|\text{b.c.}) \Bigg|_{L \rightarrow 0} = \mathcal{\widetilde{I}}^{D-1,1}_\rho (M;\beta,a_0) + \{??\},
    \end{equation}
\end{itemize}

With this discussion made evident, let us now study each possibility. Bosonic fields are treated in Sec.~\ref{Sec:Boson:P}, Sec.~\ref{Sec:Boson:DN} and Sec.~\ref{Sec:Boson:A}, while fermionic fields are considered in Sec.~\ref{Sec:Fermion:P}, Sec.~\ref{Sec:Fermion:DN} and Sec.~\ref{Sec:Fermion:A}.

%%%%%%%%%%%%%%%%%%%%%%%%%%%%%%%%%%%%%%%%%%%%%%%%%%%%%%%%%%%%%%%
%%%%%%%%%%%%%%%%%%%%%%%%%%%%%%%%%%%%%%%%%%%%%%%%%%%%%%%%%%%%%%%
%%%%%%%%%%%%%%%%%%%%%%%%%%%%%%%%%%%%%%%%%%%%%%%%%%%%%%%%%%%%%%%

\subsection{Bosonic field : periodic boundary conditions in space \label{Sec:Boson:P}}

This first case was the object of study in a previous article where we explored the subject in further detail~\cite{Cavalcanti:2018pgi}. We take the case of periodic boundary conditions, Eq.~\eqref{Eq:FuncW_Periodic_Initial}, for $a_0=0$, that is related to bosons, apply the integral representation Eq.~\eqref{Eq:K_complex} and use the following analytic extension~\cite{Elizalde:2012zza} of infinite double sum
\begin{multline}
\sum_{n_0=1,n_1=1}^\infty \frac{1}{(n_0^2 \beta^2 + n_1^2 L^2)^t} =
- \frac{\zeta(2t) }{2L^{2t}} 
+ \frac{\sqrt{\pi}}{2} \frac{\Gamma(t-1/2)}{\Gamma(t)} \frac{\zeta(2t-1)}{\beta L^{2t-1}}
\\+ \frac{2 \pi^t}{\Gamma(t)}\sqrt{\frac{L}{\beta}} \frac{1}{(\beta L)^t} \sum_{n_0,n_1=1}^{\infty} \left(\frac{n_0}{n_1}\right)^{t-\frac{1}{2}} K_{t-\frac{1}{2}} \left(2 \pi n_0 n_1 \frac{L}{\beta}\right), \label{Eq:DoubleSum_00}
\end{multline}
\noindent where $\zeta$ is the Riemann zeta function. By convention, we first do the sum over $n_0$ and then the sum over $n_1$. After this the function $\mathcal{W}_\nu$ reads
\begin{multline}
\mathcal{W}_\nu (M;\beta,a_0=0;L|\mathcal{P}) = 
\int_{c-i\infty}^{c+i\infty} \frac{dt}{4\pi i} \Gamma(t) \zeta(2t) \Gamma(t-\nu) \left(\frac{M \beta}{2}\right)^{-2t}
+ \frac{\sqrt{\pi}L}{\beta} \int_{c-i\infty}^{c+i\infty} \frac{dt}{4\pi i} \Gamma(t-\nu) \Gamma\left(t-\frac{1}{2}\right) \zeta(2t-1) \left(\frac{ML}{2}\right)^{-2t}\\
+ \frac{1}{\sqrt{\pi}} \sum_{k=1}^{\infty} \int_{c-i\infty}^{c+i\infty} \frac{ds}{2\pi i} \left(\frac{M \beta}{2\pi}\right)^{2k-2\nu} \left(\frac{\pi L}{\beta}\right)^{-2s} \Gamma(s) \zeta(2s) \Gamma\left(s+k-\nu+\frac{1}{2}\right) \zeta(2s+2k-2\nu+1).
\label{Eq:FuncW_Periodic0_Complex}
\end{multline}

A detailed treatment demands to investigate Eq.~\eqref{Eq:FuncW_Periodic0_Complex} for each different value assumed by $2\nu$ (odd, even, noninteger), as this determines whether we are dealing with single or double poles. However, motivated by previous results and to make the notation clear we choose here to exhibit only the position of the poles and the power dependencies on $\beta$ and $L$. Note that a structure as $\Gamma(u)\zeta(2u)$ means the existence of poles at $u=0,1/2$ and a structure as $\Gamma(u)\eta(2u)$ means only a pole at $u=0$. The analysis of Eq.~\eqref{Eq:FuncW_Periodic0_Complex} gives that:
\begin{itemize}
    \item for the first integral we have poles at $t=0$, $t=1/2$ and $t=\nu-j$ with $j\in[0,\infty[$. This corresponds to the dependencies $\beta^0$, $\beta^{-1}$ and $\beta^{2k-2\nu}$;
    \item for the second integral there are poles at $t=1/2$, $t=1$ and $t=\nu-j$ with $j\in[0,\infty[$. This corresponds to the dependecies $\beta^{-1}$, $\beta^{-1} L^{-1}$ and $(L/\beta) L^{2k-2\nu}$;
    \item the last integral has poles at $s=0$, $s=1/2$, $s=\nu-k-1/2$ and $s=\nu-k$. This corresponds to the dependecies $\beta^{2k-2\nu}$, $(\beta/L) \beta^{2k-2\nu}$, $L^{2k-2\nu}$ and $(L/\beta) L^{2k-2\nu}$.
\end{itemize}
    
We are mainly interested in the behavior of $\mathcal{I}^{D,2}$ as $L\rightarrow 0$ to see whether there is some function of the inverse temperature $\beta=1/T$ that could be related to a scenario with one less dimension. To do this we use some \textit{scale} function multiplied by the Feynman amplitude,
\begin{equation}
s(L) \mathcal{I}^{D,2}_\rho (M;\beta,a_0=0;L|\mathcal{P}) \Bigg|_{L \rightarrow 0}.
\end{equation}

In a previous article we used $s(L)=L$ and split this product into three different parts: one that goes to zero as $L\rightarrow0$ and therefore do not contribute in anything, another component that grows as $L\rightarrow0$ and could be considered a residual contribution coming from high dimension and a final component that gives a contribution independent of the length $L$. From the analysis of the poles and the power dependencies on $\beta$ and $L$ we can note that the relevant poles are $t=1$ from the second integral and $s=1/2$ from the third integral in Eq.~\eqref{Eq:FuncW_Periodic0_Complex}. Indeed, this gives the simple result
\begin{equation}
L \mathcal{I}^{D,2}_\rho (M;\beta,a_0=0;L|\mathcal{P}) \Bigg|_{L \rightarrow 0} = \mathcal{I}^{D-1,1}_\rho (M;\beta,a_0=0) + \text{divergent terms}.
\label{Eq:DimRed_BosonPeriodic}
\end{equation}

Where $\mathcal{I}^{D-1,1}_\rho (M;\beta,a_0=0)$ is exactly the Feynamn amplitude for the reduced scenario with $D-1$ dimensions and just one compactification related to the temperature. It reads,
\begin{equation*}
\mathcal{I}^{D-1,1}_\rho (M;\beta,a_0=0) = \frac{\left(M^2/2\right)^\nu}{(2\pi)^\frac{D}{2}2^{\rho-2}\Gamma(\rho)}    
\left[
\frac{\pi}{M^2\beta} \Gamma(1-\nu) 
+ \frac{\sqrt{\pi}}{M} \sum_{k=0}^\infty \frac{(-1)^k}{\Gamma(k+1)} \Gamma\left(\nu-k-\frac{1}{2}\right) \zeta(2\nu-2k-1) \left(\frac{M \beta}{2}\right)^{-2\nu+2k+1}
\right].
\end{equation*}

This result shows that the dimensional reduction is well-defined for a self-interacting bosonic field with periodic boundary conditions, as already discussed in the previous article. For further details, one is referred to Ref.~\cite{Cavalcanti:2018pgi} where this relation was obtained with a careful investigation for even, odd and noninteger $D$ and also the residual divergent terms were fully exhibited. The important aspect to be noted here is that we can get the structure of the function from a quick investigation of the poles. To avoid a lengthy exposition, this procedure is repeated in the following sections to study other cases of interest.

%%%%%%%%%%%%%%%%%%%%%%%%%%%%%%%%%%%%%%%%%%%%%%%%%%%%%%%%%%%%%%%
%%%%%%%%%%%%%%%%%%%%%%%%%%%%%%%%%%%%%%%%%%%%%%%%%%%%%%%%%%%%%%%
%%%%%%%%%%%%%%%%%%%%%%%%%%%%%%%%%%%%%%%%%%%%%%%%%%%%%%%%%%%%%%%

\subsection{Bosonic field : Dirichlet and Neumann boundary conditions in space \label{Sec:Boson:DN}}

As discussed previously, both the Dirichlet (Eq.~\eqref{Eq:FuncI_Dirichlet}) and Neumann (Eq.~\eqref{Eq:FuncI_Neumann}) boundary conditions are a linear combination of a model with periodic boudary condition in space and a dimensionally reduced model. Therefore, as we know that the behavior of the model with periodic boundary conditions in space is given by Eq.~\eqref{Eq:DimRed_BosonPeriodic}, we obtain directly that
\begin{align}
L \mathcal{I}^{D,2}_\rho (M;\beta,a_0=0;L|\mathcal{D}) \Bigg|_{L \rightarrow 0} &= -\frac{1}{4} \mathcal{I}^{D-1,1}_\rho (M;\beta,0) + \text{divergent terms},
\label{Eq:DimRed_BosonDirichlet}\\
L \mathcal{I}^{D,2}_\rho (M;\beta,a_0=0;L|\mathcal{N}) \Bigg|_{L \rightarrow 0} &= \frac{3}{4} \mathcal{I}^{D-1,1}_\rho (M;\beta,0) + \text{divergent terms},
\label{Eq:DimRed_BosonNeumann}
\end{align}

Just like the scenario with periodic boundary conditions, we obtain that the dimensional reduction is well-defined. What changes is the relation between the ($D$)-dimensional model and the ($D-1$)-dimensional model. The significance of this can be further understood if we follow the discussion of a previous article~\cite{Cavalcanti:2018pgi} and consider a bosonic model with quartic interaction given by the coupling constant $\lambda_D$. The  relationship between the coupling constant of the dimensionally reduced model $\lambda_{D-1}$ and $\lambda_D$ is different for each boundary condition,
\begin{align*}
\lambda_{D-1} = \frac{\lambda_D}{L},& \quad \text{Periodic b.c.};\\
\lambda_{D-1} = -\frac{\lambda_D}{4L},& \quad \text{Dirichlet b.c.};\\
\lambda_{D-1} = \frac{3\lambda_D}{4L},& \quad \text{Neumann b.c.}.
\end{align*}
Notice that for Dirichlet boundary conditions the coupling constant of the dimensionally reduced model changes sign, which raises a question about the vacua stability of this model and motivates a further investigation. 

%%%%%%%%%%%%%%%%%%%%%%%%%%%%%%%%%%%%%%%%%%%%%%%%%%%%%%%%%%%%%%%
%%%%%%%%%%%%%%%%%%%%%%%%%%%%%%%%%%%%%%%%%%%%%%%%%%%%%%%%%%%%%%%
%%%%%%%%%%%%%%%%%%%%%%%%%%%%%%%%%%%%%%%%%%%%%%%%%%%%%%%%%%%%%%%

\subsection{Bosonic field : antiperiodic boundary conditions in space \label{Sec:Boson:A}}

In this section we consider bosonic fields ($a_0=0$) with antiperiodic boundary conditions in space. To investigate this, we apply the integral representation of the $K_\nu$, Eq.~\eqref{Eq:K_complex}, in the function $\mathcal{W}_\nu$, Eq.~\eqref{Eq:FuncW_AntiPeriodic_Initial}, and make use of the analytic extension that reads
\begin{multline}
\sum_{n_0=1,n_1=1}^\infty \frac{(-1)^{n_1}}{(n_0^2 \beta^2 + n_1^2 L^2)^t} =
\frac{1}{2L^{2t}} \eta(2t) 
- \frac{\sqrt{\pi}}{2} \frac{\Gamma(t-1/2)}{\Gamma(t)} \frac{\eta(2t-1)}{\beta L^{2t-1}}
\\+ \frac{2 \pi^t}{\Gamma(t)}\sqrt{\frac{L}{\beta}} \frac{1}{(\beta L)^t} \sum_{n_0,n_1=1}^{\infty} (-1)^{n_1} \left(\frac{n_0}{n_1}\right)^{t-\frac{1}{2}} K_{t-\frac{1}{2}} \left(2 \pi n_0 n_1 \frac{L}{\beta}\right),\label{Eq:DoubleSum_10}
\end{multline}
\noindent where $\eta$ is the Dirichlet eta function. By convention, we first do the sum over $n_0$ and then the sum over $n_1$. After this, The function $\mathcal{W}_\nu$ reads
\begin{multline}
\mathcal{W}_\nu (M;\beta,a_0=0;L|\mathcal{A}) = 
\int_{c-i\infty}^{c+i\infty} \frac{dt}{4\pi i} \Gamma(t) \zeta(2t) \Gamma(t-\nu) \left(\frac{M \beta}{2}\right)^{-2t}
- \frac{\sqrt{\pi}L}{\beta} \int_{c-i\infty}^{c+i\infty} \frac{dt}{4\pi i} \Gamma(t-\nu) \Gamma\left(t-\frac{1}{2}\right) \eta(2t-1) \left(\frac{ML}{2}\right)^{-2t}\\
- \frac{1}{\sqrt{\pi}} \sum_{k=1}^{\infty} \int_{c-i\infty}^{c+i\infty} \frac{ds}{2\pi i} \left(\frac{M \beta}{2\pi}\right)^{2k-2\nu} \left(\frac{\pi L}{\beta}\right)^{-2s} \Gamma(s) \eta(2s) \Gamma\left(s+k-\nu+\frac{1}{2}\right) \zeta(2s+2k-2\nu+1).
\label{Eq:FuncW_Antiperiodic0_Complex}
\end{multline}

We investigate the above equation and obtain the poles for each of the integrals. 
\begin{itemize}
    \item First integral: poles at $t=0$, $t=1/2$ and $t=\nu-j$ with $j\in[0,\infty[$. This corresponds to the dependencies $\beta^0$, $\beta^{-1}$ and $\beta^{2k-2\nu}$.
    \item Second integral: poles at $t=1/2$ and $t=\nu-j$ with $j\in[0,\infty[$. This corresponds to the dependecies $\beta^{-1}$ and $(L/\beta) L^{2k-2\nu}$.
    \item Third integral: poles at $s=0$, $s=\nu-k-1/2$ and $s=\nu-k$. This corresponds to the dependecies $\beta^{2k-2\nu}$, $L^{2k-2\nu}$ and $(L/\beta) L^{2k-2\nu}$.
\end{itemize}

This means that the case of antiperiodic boundary conditions in space and $a_0=0$ only has dependencies as $\beta^\alpha, (L/\beta) L^\alpha, L^\alpha$. Therefore, the procedure of taking $L\rightarrow 0$,
\begin{equation}
L \mathcal{I}^{D,2}_\rho (M;\beta,a_0=0;L|\mathcal{A}) \Bigg|_{L \rightarrow 0},
\end{equation}
\noindent does not reproduce any behavior of a model with fewer dimensions. This is completely different from the situation with periodic boundary conditions in space, where a relationship between a ``film" model ($D$ dimensions) and a ``surface" model ($D-1$ dimensions)  is clear. Therefore, for a bosonic model with antiperiodic boundary condition in space, the idea of dimensional reduction is ill-defined and does not result in any temperature-dependent function.

%%%%%%%%%%%%%%%%%%%%%%%%%%%%%%%%%%%%%%%%%%%%%%%%%%%%%%%%%%%%%%%
%%%%%%%%%%%%%%%%%%%%%%%%%%%%%%%%%%%%%%%%%%%%%%%%%%%%%%%%%%%%%%%
%%%%%%%%%%%%%%%%%%%%%%%%%%%%%%%%%%%%%%%%%%%%%%%%%%%%%%%%%%%%%%%

\subsection{Fermionic field : periodic boundary conditions in space \label{Sec:Fermion:P}}

From this point forward we proceed to take into account the situation of a fermionic model. We already know that the one loop Feynman amplitude for fermions is related to the one loop Feynman amplitude for bosons with $a_0=1$, this relation is given by Eq.~\eqref{Eq:RelationJ_I}. At first, we consider periodic boundary conditions in space, given by Eq.~\eqref{Eq:FuncW_Periodic_Initial}. To explore the behavior as $L\rightarrow0$ we use the integral representation of $K_\nu$, Eq.~\eqref{Eq:K_complex}, and the double sum that arises is treated by an analytic extension,
\begin{multline}
\sum_{n_0=1,n_1=1}^\infty \frac{(-1)^{n_0}}{(n_0^2 \beta^2 + n_1^2 L^2)^t} = 
- \frac{1}{2L^{2t}} \zeta(2t) 
+ \frac{2 \pi^t}{\Gamma(t)}\sqrt{\frac{L}{\beta}} \frac{1}{(\beta L)^t} \sum_{n_0,n_1=1}^{\infty} \left(\frac{n_0}{n_1}\right)^{t-\frac{1}{2}} \left[ - K_{t-\frac{1}{2}} \left(2 \pi n_0 n_1 \frac{L}{\beta}\right) \right.\\\left. + 2^{\frac{1}{2}-t} K_{t-\frac{1}{2}} \left(2 \pi n_0 n_1 \frac{L}{\beta}\right)\right].\label{Eq:DoubleSum_01}
\end{multline}
\noindent Hence, the function $\mathcal{W}_\nu$ reads
    \begin{multline}
    \mathcal{W}_\nu (M;\beta,a_0=1;L|\mathcal{P}) = 
    - \int_{c-i\infty}^{c+i\infty} \frac{dt}{4\pi i} \Gamma(t) \eta(2t) \Gamma(t-\nu) \left(\frac{M \beta}{2}\right)^{-2t}
    \\
    \frac{1}{\sqrt{\pi}} \sum_{k=1}^{\infty} \int_{c-i\infty}^{c+i\infty} \frac{ds}{2\pi i} \left(\frac{M \beta}{2\pi}\right)^{2k-2\nu} \left(\frac{\pi L}{\beta}\right)^{-2s} \Gamma(s) \zeta(2s) \Gamma\left(s+k-\nu+\frac{1}{2}\right) \left(-1 + 2^{2s+2k-2\nu+1}\right) \zeta(2s+2k-2\nu+1),
    \label{Eq:FuncW_Periodic1_Complex}
    \end{multline}
    \noindent and an analysis of each term gives that
    \begin{itemize}
        \item for the first integral there are poles at $t=0$ and $t=\nu-j$ with $j\in[0,\infty[$. This corresponds to the dependencies $\beta^0$ and $\beta^{2k-2\nu}$;
        \item and for the second integral there are poles at $s=0$, $s=1/2$ and $s=\nu-k$. This corresponds to the dependecies $\beta^{2k-2\nu}$, $(\beta/L) \beta^{2k-2\nu}$ and $(L/\beta) L^{2k-2\nu}$.
    \end{itemize}    

It can be noted that the relevant contribution comes from the pole $s=1/2$ of the second integral. This is the contribution that survives at $L\rightarrow0$. Making it explicit, we obtain in this limit that
\begin{equation}
L \mathcal{I}^{D,2}_\rho (M;\beta,a_0=1;L|\mathcal{P}) \Bigg|_{L \rightarrow 0} = \mathcal{F}^D_\rho(M,\beta;c_1=4,c_2=0) + \text{divergent terms},
\label{Eq:DimRed_FermionPeriodic}
\end{equation}
\noindent where the function $\mathcal{F}^D_\rho(M,\beta;c_1,c_2)$ is defined in Eq.~\eqref{Eq:F_function}.

Just as we did when we exhibited the result for the bosonic case ($a_0=0$) in periodic boundary conditions in space, let us concentrate on the behavior as $L\rightarrow0$. To make the comparison clear, we can keep in mind the analogy of heated ``films" (in dimension $D$ with two compactifications) and ``surfaces" (in dimension $D-1$ with one compactification). The heated ``film" described by a fermionic model is given by \eqref{Eq:DimRed_FermionPeriodic} when the film thickness is very small. However, the ``surface" described by the same fermionic model reads
\begin{equation}
\mathcal{I}^{D-1,1}_\rho (M;\beta,a_0=1) = \mathcal{F}^D_\rho(M,\beta;c_1=0,c_2=1/2) + \text{divergent terms},
\end{equation}
\noindent which is completely different.

Therefore, in the case of a fermionic model, there is no direct relationship between models in different dimensions. This result resembles the discussion that the procedure of dimensional reduction and quantization does not commute for fermionic models~\cite{Ospedal:2017ubh} and that the dimensional reduction behaves differently for bosons and fermions~\cite{Huang:1995um}. 

%%%%%%%%%%%%%%%%%%%%%%%%%%%%%%%%%%%%%%%%%%%%%%%%%%%%%%%%%%%%%%%
%%%%%%%%%%%%%%%%%%%%%%%%%%%%%%%%%%%%%%%%%%%%%%%%%%%%%%%%%%%%%%%
%%%%%%%%%%%%%%%%%%%%%%%%%%%%%%%%%%%%%%%%%%%%%%%%%%%%%%%%%%%%%%%

\subsection{Fermionic field : Dirichlet and Neumann boundary conditions in space \label{Sec:Fermion:DN}}

As a next step, we investigate the fermionic field at different spatial boundary conditions. Just as done in Sec.~\ref{Sec:Boson:DN} for bosonic fields in Dirichlet and Neumann boundary conditions, we apply in Eq.~\eqref{Eq:FuncI_Dirichlet} and Eq.~\eqref{Eq:FuncI_Neumann} the known result for periodic boundary conditions, Eq.~\eqref{Eq:DimRed_FermionPeriodic}, and the dimensionally reduced fermionic model given by Eq.~\eqref{Eq:Id1_fermion}. This gives, respectively, for Dirichlet and Neumann boundary conditions that
\begin{align}
L \mathcal{I}^{D,2}_\rho (M;\beta,a_0=1;L|\mathcal{D}) \Bigg|_{L \rightarrow 0} &= \frac{3}{4} \mathcal{F}^D_\rho(M,\beta;c_1=4/3,c_2=1/3) + \text{divergent terms},
\label{Eq:DimRed_FermionDirichlet}\\
L \mathcal{I}^{D,2}_\rho (M;\beta,a_0=1;L|\mathcal{N}) \Bigg|_{L \rightarrow 0} &= -\frac{1}{4} \mathcal{F}^D_\rho(M,\beta;c_1=-4,c_2=1) + \text{divergent terms}.
\label{Eq:DimRed_FermionNeumann}
\end{align}

These results reinforce that, as found in Sec.~\ref{Sec:Fermion:P}, the fermionic field does not undergo a dimensional reduction as bosonic fields. We can, indeed, obtain a dimensionally reduced model, as expressed in Eq.~\eqref{Eq:DimRed_FermionDirichlet} and Eq.~\eqref{Eq:DimRed_FermionNeumann}. However, it has no relation with the otherwise expected result given by Eq.~\eqref{Eq:Id1_fermion}.

%%%%%%%%%%%%%%%%%%%%%%%%%%%%%%%%%%%%%%%%%%%%%%%%%%%%%%%%%%%%%%%
%%%%%%%%%%%%%%%%%%%%%%%%%%%%%%%%%%%%%%%%%%%%%%%%%%%%%%%%%%%%%%%
%%%%%%%%%%%%%%%%%%%%%%%%%%%%%%%%%%%%%%%%%%%%%%%%%%%%%%%%%%%%%%%

\subsection{Fermionic field : antiperiodic boundary conditions in space \label{Sec:Fermion:A}}

At last, let us consider a fermionic model ($a_0=1$) with antiperiodic boundary conditions in space (Eq.~\eqref{Eq:FuncW_AntiPeriodic_Initial}). After using the integral representation of Eq.~\eqref{Eq:K_complex} we use the following analytic extension of the double sum,
    \begin{multline}
    \sum_{n_0=1,n_1=1}^\infty \frac{(-1)^{n_0+n_1}}{(n_0^2 \beta^2 + n_1^2 L^2)^t} =
    \frac{1}{2L^{2t}} \eta(2t) 
    + \frac{2 \pi^t}{\Gamma(t)}\sqrt{\frac{L}{\beta}} \frac{1}{(\beta L)^t} \sum_{n_0,n_1=1}^{\infty} (-1)^{n_1} \left(\frac{n_0}{n_1}\right)^{t-\frac{1}{2}} \left[ - K_{t-\frac{1}{2}} \left(2 \pi n_0 n_1 \frac{L}{\beta}\right) \right.\\\left. + 2^{\frac{1}{2}-t} K_{t-\frac{1}{2}} \left(2 \pi n_0 n_1 \frac{L}{\beta}\right)\right], \label{Eq:DoubleSum_11}
    \end{multline}
\noindent and obtain the expression for the function $\mathcal{W}_\nu$
    \begin{multline}
    \mathcal{W}_\nu (M;\beta,a_0=1;L|\mathcal{A}) = 
    - \int_{c-i\infty}^{c+i\infty} \frac{dt}{4\pi i} \Gamma(t) \eta(2t) \Gamma(t-\nu) \left(\frac{M \beta}{2}\right)^{-2t}
    \\
    - \frac{1}{\sqrt{\pi}} \sum_{k=1}^{\infty} \int_{c-i\infty}^{c+i\infty} \frac{ds}{2\pi i} \left(\frac{M \beta}{2\pi}\right)^{2k-2\nu} \left(\frac{\pi L}{\beta}\right)^{-2s} \Gamma(s) \eta(2s) \Gamma\left(s+k-\nu+\frac{1}{2}\right) \left(-1 + 2^{2s+2k-2\nu+1}\right) \zeta(2s+2k-2\nu+1).
    \label{Eq:FuncW_Antiperiodic1_Complex}
    \end{multline}

    Studying the poles for each integral in Eq.~\eqref{Eq:FuncW_Antiperiodic1_Complex} we obtain that
    \begin{itemize}
        \item First integral: poles at $t=0$ and $t=\nu-j$ with $j\in[0,\infty[$. This corresponds to the dependencies $\beta^0$ and $\beta^{2k-2\nu}$;
        \item Second integral: poles at $s=0$ and $s=\nu-k$. This corresponds to the dependecies $\beta^{2k-2\nu}$ and $L^{2k-2\nu}$.
    \end{itemize}
    Therefore, for antiperiodic boundary conditions in space and $a_0=1$ there is no mixed dependency on $\beta$ and $L$. Also, just like the case of antiperiodic boundary conditions in space for bosons discussed in Sec.~\ref{Sec:Boson:A} the procedure of dimensional reduction is ill-defined.
    
    This result shows that the use of antiperiodic boundary conditions in space forbids the procedure of dimensional reduction both for bosonic and fermionic fields. This might be an indication of a topological aspect, independent of the nature of the field.

%%%%%%%%%%%%%%%%%%%%%%%%%%%%%%%%%%%%%%%%%%%%%%%%%%%%%%%%%%%%%%%
%%%%%%%%%%%%%%%%%%%%%%%%%%%%%%%%%%%%%%%%%%%%%%%%%%%%%%%%%%%%%%%
%%%%%%%%%%%%%%%%%%%%%%%%%%%%%%%%%%%%%%%%%%%%%%%%%%%%%%%%%%%%%%%

\section{Conclusion}

We discussed in Sec.~\ref{Sec:DimRed} that there were three possible outcomes when one investigates the procedure of dimensional reduction as proposed in this article. In the remaining sections we found examples of all three categories:
\begin{itemize}
    \item Well-defined dimensional reduction.
    
    This happens for bosonic fields in periodic, Dirichlet and Neumann boundary conditions where there is a simple relation between a model in $D$ dimensions that is dimensionally reduced and a model in $D-1$ dimensions. See Sec.~\ref{Sec:Boson:P} and Sec.~\ref{Sec:Boson:DN}.
    \item Ill-defined dimensional reduction.
    
    This happens for antiperiodic boundary conditions in space, both for bosonic and fermionic fields. See Sec.~\ref{Sec:Boson:A} and Sec.~\ref{Sec:Fermion:A}.
    \item Dimensional reduction to a different model.
    
    This happens for fermionic fields in periodic, Dirichlet and Neumann boundary conditions where the model in $D$ dimensions that is dimensionally reduced has no relation with a model originally constructed in $D-1$ dimensions. See Sec.~\ref{Sec:Fermion:P} and Sec.~\ref{Sec:Fermion:DN}.
\end{itemize}

We found that the previous article~\cite{Cavalcanti:2018pgi} was indeed a special case (bosonic field, periodic boundary condition in space) and now we exhibit a bigger picture of the problem. The procedure of dimensional reduction indeed depends on the imposed boundary conditions and the nature of the field. Nevertheless, there are yet some open questions. The behavior of fermionic fields passing through a dimensional reduction might be explained by the fact that fermions are dependent on the number of spatial dimensions. Moreover, the forbidden dimensional reduction for models with antiperiodic boundary conditions in space is perhaps a topological aspect of dimensionally reducing a M\"obius strip, which would explain the independence on the nature of the fields.

%%%%%%%%%%%%%%%%%%%%%%%%%%%%%%%%%%%%%%%%%%%%%%%%%%%%%%%%%%%%%%%
%%%%%%%%%%%%%%%%%%%%%%%%%%%%%%%%%%%%%%%%%%%%%%%%%%%%%%%%%%%%%%%
%%%%%%%%%%%%%%%%%%%%%%%%%%%%%%%%%%%%%%%%%%%%%%%%%%%%%%%%%%%%%%%

\acknowledgments{The authors thank the Brazilian agency Conselho Nacional de Desenvolvimento Cient\'ifico e Tecnol\'ogico (CNPq) for partial financial support.}

\appendix

%%%%%%%%%%%%%%%%%%%%%%%%%%%%%%%%%%%%%%%%%%%%%%%%%%%%%%%%%%%%%%%
%%%%%%%%%%%%%%%%%%%%%%%%%%%%%%%%%%%%%%%%%%%%%%%%%%%%%%%%%%%%%%%
%%%%%%%%%%%%%%%%%%%%%%%%%%%%%%%%%%%%%%%%%%%%%%%%%%%%%%%%%%%%%%%

\section{Relation between fermionic and bosonic integrals\label{Ap:Fermion_Boson}}

The one-loop Feynman amplitude for self interacting fermionic field with coupling $a+b\gamma_S$ is
\begin{equation}
\mathcal{J}_\nu^D(m) = \text{tr}\int \frac{d^D p}{(2\pi)^D} \left(\frac{a+b\gamma_S}{i \slashed{p} + m}\right)^\nu = \text{tr}\int \frac{d^D p}{(2\pi)^D} \left(\frac{(a+b\gamma_S)(-i \slashed{p} + m)}{p^2 + m^2}\right)^\nu.
\end{equation}
\noindent Here we use the notation of Ref~\cite{ZinnJustin:2002ru} for the Euclidean Dirac matrices. To compute the trace in a systematic way we define $v = -i p^\mu \gamma_\mu + m$, $\widetilde{v} = i p^\mu \gamma_\mu + m$ and note that $v \gamma_S = \widetilde{v} \gamma_S$. Organizing the trace $\mathcal{T}_\nu = \text{tr} \left[(a+b\gamma_S)v\right]^\nu$ in such a way that all $\gamma_S$ matrices are on the left we have
\begin{subequations}
    \begin{align}
    \mathcal{T}_1(a,b) &= a v + b \gamma_S v,\\
    \mathcal{T}_2(a,b) &= a^2 v^2 + a b \gamma_S (\widetilde{v} v + v^2) + b^2 \gamma_S^2 \widetilde{v} v,\\
    \mathcal{T}_3(a,b) &= a^3 v^3 + a^2 b \gamma_S (\widetilde{v}^2 v + \widetilde{v} v^2 + v^3) + a b^2 \gamma_S^2 (\widetilde{v}^2 v + 2\widetilde{v} v^2)+ b^3 \gamma_S^3 \widetilde{v} v^2,\\
    \mathcal{T}_4(a,b) &= a^4 v^4 + a^3 b \gamma_S (\widetilde{v}^3 v + \widetilde{v}^2 v^2 + \widetilde{v} v^3 + v^4) + a^2 b^2 \gamma_S^2 (\widetilde{v}^3 v + 2 \widetilde{v}^2 v^2 + 3 \widetilde{v} v^3)+ a b^3 \gamma_S^3 (2 \widetilde{v}^2 v^2 + 2 \widetilde{v} v^3) + b^4 \gamma_S^4 \widetilde{v}^2 v^2.
    \end{align}
\end{subequations}

From these we can infer some relations regarding the trace $\mathcal{T}_\nu$ for any $\nu$. The component with $b=0$ and $a\neq 0$ contributes as
\begin{equation*}
\mathcal{T}_\nu(a,0) = a^\nu v^\nu,
\end{equation*}
\noindent and the component with $a=0$ and $b\neq 0$ behaves as
\begin{equation*}
\mathcal{T}_\nu(0,b) = b^\nu \gamma_S^\nu (\widetilde{v}v)^{\floor{\nu/2}} v^{\nu - 2 \floor{\nu/2}}.
\end{equation*}

The mixed terms are a little bit more intricated. First, we adopt another notation defining some function $f_j^{(i)} (\widetilde{v}, v)$,
\begin{equation*}
\mathcal{T}_\nu(a,b) - \mathcal{T}_\nu(a,0) - \mathcal{T}_\nu(0,b) = \sum_{\sigma=1}^{\nu-1} a^{\nu-\sigma} b^\sigma \gamma_S^\sigma f_\sigma^{(\nu-\sigma)} (\widetilde{v}, v),
\end{equation*}
\noindent where the function $f_j^{(i)} (\widetilde{v}, v)$ can be shown to satisfy the following difference equations
\begin{align}
f_{2\ell}^{(i)} (\widetilde{v}, v) &= \frac{\widetilde{v} v}{\ell!} \frac{\partial}{\partial v} f_{2\ell-1}^{(i)} (\widetilde{v}, v),\\
f_{2\ell+1}^{(i)} (\widetilde{v}, v) &= \frac{\widetilde{v} v}{\ell!} \frac{\partial}{\partial \widetilde{v}} f_{2\ell}^{(i)} (\widetilde{v}, v).
\end{align}

Therefore, once we obtain one of these functions all others are obtained recursively. The simpler one is the case $f_1^{(i)}$ which is associated with $a^i b \gamma_S$ and can be directly written as
\begin{equation*}
f_1^{(i)} = \sum_{k=0}^{i} \widetilde{v}^{i-k} v^{k+1}.
\end{equation*}

With this in hand we use the difference equations and obtain the generalization that
\begin{equation*}
f_j^{(i)} (\widetilde{v}, v) = \sum_{k=0}^{i} \frac{(k+\floor{j/2})!}{k!\floor{j/2}!} \frac{(i-k+\floor{\frac{j+1}{2}}-1)!}{(i-k)!(\floor{\frac{j+1}{2}}-1)!} \widetilde{v} ^{i+j-k-\floor{\frac{j+1}{2}}} v^{k + \floor{\frac{j+1}{2}}}
\end{equation*}

Substituting back, we obtain that the complete trace is
\begin{multline}
\text{tr} \left[(a+b\gamma_S)v\right]^\nu = \text{tr} \Bigg[
a^\nu v^\nu 
+ b^\nu \gamma_S^\nu (p^2+m^2)^{\floor{\frac{\nu}{2}}} v^{\nu-2\floor{\frac{\nu}{2}}}
\\+ \sum_{\sigma=1}^{\nu-1} a^{\nu-\sigma} b^\sigma \gamma_S^\sigma \sum_{k=0}^{\nu-\sigma} \frac{\left(k+\floor{\frac{\sigma}{2}}\right)!}{k! \floor{\frac{\sigma}{2}}!} \frac{\left(\nu-k-1-\floor{\frac{\sigma}{2}}\right)!}{\left(\nu-k-\sigma\right)!\left(\floor{\frac{\sigma+1}{2}}-1\right)!} \bar{v}^{\nu-k-\floor{\frac{\sigma+1}{2}}} v^{k+\floor{\frac{\sigma+1}{2}}}
\Bigg].
\end{multline}

Therefore, the trace operation becomes simply
\begin{equation*}
\text{tr} (\widetilde{v} v)^n (v^2)^m = \text{tr} (\widetilde{v} v)^n (\widetilde{v}^2)^m =d_\gamma (m^2+p^2)^n (m^2-p^2)^m,
\end{equation*}
\noindent where $d_\gamma$ is the dimension of the gamma matrix.

After computing the full trace and making some algebraic manipulation we obtain
\begin{multline}
\frac{1}{d_\gamma} \text{tr} \left[(a+b\gamma_S)v\right]^\nu  =
a^\nu \sum_{k=0}^{\floor{\frac{\nu}{2}}} \binom{\nu}{2k} m^{\nu-2k} (-p^2)^{k} 
+ b^\nu (p^2+m^2)^{\frac{\nu}{2}} \left(\nu - 2 \floor{\nu/2}\right)
\\
+ \sum_{k=1}^{\floor{\frac{\nu-1}{2}}} a^{\nu-2k} b^{2k} \sum_{j=k}^{\floor{\frac{\nu}{2}}} \frac{j! (\nu-j-1)!}{(j-k)!k!(\nu-k-j)!(k-j)!} (p^2+m^2)^j \sum_{\ell=0}^{\floor{\frac{\nu}{2}-j}} \binom{\nu-2j}{2\ell} m^{\nu-2j-2\ell} (-p^2)^\ell
\\
+ \sum_{k=1}^{\floor{\frac{\nu-1}{2}}} a^{\nu-2k} b^{2k} \sum_{j=\floor{\frac{\nu}{2}}+1}^{\nu-k} \frac{j! (\nu-j-1)!}{(j-k)!k!(\nu-k-j)!(k-j)!} (p^2+m^2)^{\nu-j} \sum_{\ell=0}^{\floor{\frac{\nu}{2}-j}} \binom{2j-\nu}{2\ell} m^{2j-\nu-2\ell} (-p^2)^\ell.
\end{multline}
\noindent As a final manipulation we use that
\begin{equation*}
(-p^2)^\ell = \sum_{n=0}^{\ell} (-1)^n \binom{\ell}{n} (p^2+m^2)^n m^{2\ell-2n},
\end{equation*}
\noindent and now we can relate the fermionic scenario with the bosonic one,
\begin{multline*}
\frac{1}{d_\gamma} \mathcal{J}_\nu^D  =
a^\nu \sum_{k=0}^{\floor{\frac{\nu}{2}}} \sum_{j=0}^{k} \binom{\nu}{2k} \binom{k}{j} m^{\nu-2j} (-1)^j \mathcal{I}^D_{\nu-j} (m^2)
+ b^\nu \left(\nu - 2 \floor{\nu/2}\right) \mathcal{I}^D_{\nu/2} (m^2)
\\
+ \sum_{k=1}^{\floor{\frac{\nu-1}{2}}} a^{\nu-2k} b^{2k} \sum_{j=k}^{\floor{\frac{\nu}{2}}} \frac{j! (\nu-j-1)!}{(j-k)!k!(\nu-k-j)!(k-j)!} \sum_{\ell=0}^{\floor{\frac{\nu}{2}-j}} \binom{\nu-2j}{2\ell}  \sum_{n=0}^{\ell} (-1)^n \binom{\ell}{n} m^{\nu-2j-2n} \mathcal{I}^D_{\nu-j-n} (m^2)
\\
+ \sum_{k=1}^{\floor{\frac{\nu-1}{2}}} a^{\nu-2k} b^{2k} \sum_{j=\floor{\frac{\nu}{2}}+1}^{\nu-k} \frac{j! (\nu-j-1)!}{(j-k)!k!(\nu-k-j)!(k-j)!} \sum_{\ell=0}^{\floor{\frac{\nu}{2}-j}} \binom{2j-\nu}{2\ell} \sum_{n=0}^{\ell} \binom{\ell}{n} m^{2j-\nu-2n} \mathcal{I}^D_{j-n} (m^2).
\end{multline*}

This relation also holds if one considers compactified dimensions. One must only be careful that the conditions imposed on $\mathcal{I}$ will be, in this case, the conditions that would be imposed on the fermionic integral. Therefore, if one introduces a compactification of the imaginary time to introduce temperature, it must have antiperiodic boundary condition as we are dealing with fermions.

\bibliography{Q_Refs}{}
\bibliographystyle{apsrev4-1}

\end{document}